\documentclass[intlimits,twoside,a4paper]{article}

\usepackage{amsmath,amssymb}
\usepackage{graphicx}

\usepackage[T2A]{fontenc}
\usepackage[cp1251]{inputenc}

\usepackage[eqsecnum]{cmpj2}

\issue{2014}{17}{2}{23301}
\doinumber{10.5488/CMP.17.23301}

\title[Conformational properties of polymers in anisotropic environments]%
{Conformational properties of polymers in anisotropic environments}
\author[K. Haydukivska, V. Blavatska]{K. Haydukivska, V. Blavatska}
\address{
Institute for Condensed Matter Physics of the  National Academy of  Sciences of Ukraine, \\ 1 Svientsitski St., 79011 Lviv, Ukraine}
\date{Received November 28, 2013, in final form April 21, 2014}

\begin{document}

\maketitle

\begin{abstract}
We analyze the conformational properties of  polymer macromolecules in solutions in presence
of extended structural obstacles of (fractal) dimension $\varepsilon_d$ causing the anisotropy of environment.
Applying the pruned-enriched Rosenbluth method (PERM),
we obtain numerical estimates for scaling exponents and universal shape parameters of
polymers in such environments for a wide range $0<\varepsilon_d<2$ in space dimension $d=3$.
An analytical description of the model is developed within the des Cloizeaux direct polymer renormalization scheme.
Both numerical and analytical studies qualitatively confirm the
existence of two characteristic length scales of polymer chain in directions
parallel and perpendicular to the extended defects.


\end{abstract}


\section{Introduction}

Many physical objects are characterized by anisotropy of structure: real magnetic crystals often contain extended defects in
the form of linear dislocations, disclinations or planar regions of different phases \cite{Dorogovtsev80,Yamazaki86,Yamazaki01,Yamazaki02,Yamazaki03};
in polymer systems,  the understanding of the behavior of macromolecules
in solutions having spatial anisotropy caused by the presence of fibrous obstacles is of great importance, e.g., in gels \cite {Stylianopoulos10}, intra- and extracellular environment \cite{Xiao08,Verkman13}, or in the vicinity of planes (membranes) \cite{Cannel80}.

The analytical description of crystalline materials  with extended defects  attracts a lot of interest
\cite{Dorogovtsev80,Yamazaki86,Yamazaki01,Yamazaki02,Yamazaki03,Boyanovsky82,Lawrie84,Lee92,Cesare94,Korzhenevskii96}.
In particular, Dorogovtsev \cite{Dorogovtsev80} proposed the model
of a $d$-dimensional spin system with quenched nonmagnetic
defects in the form of $\varepsilon_d$-dimensional objects,
which are randomly distributed over the remaining
$d-\varepsilon_d$ dimensions. The anisotropy of the system brings
about two different characteristic  length scales
(correlation lengths  $\xi_{||}$ and $\xi_{\perp}$), reflecting
the macroscopic properties of the system along the directions
``parallel'' to the $\varepsilon_d$-dimensional defect and along
the ``perpendicular'' directions:
\begin {equation}
\xi_{||}\sim |t|^{-\nu_{||}},\phantom{5555555} \xi_{\perp}\sim
|t|^{-\nu_{\perp}},\label{xi}
\end{equation}
where $t$ is the reduced distance to the
critical temperature $t=(T-T_\mathrm{c})/T_\mathrm{c}$,   $\nu_{\perp}$ and $\nu_{||}$ are universal critical exponents.

 A number of  conformational properties of long flexible polymer chains in solutions
can also be described within the critical exponent formalism:  for
example, the  averaged mean-squared  distance between two ends of a
chain obeys the scaling law:
 \begin{equation}
\langle R^2 \rangle  \sim  N^{2\nu}, \label {RR}
\end{equation}
where $N$ is the number of monomers in the chain and $\nu$ is an
universal quantity that does not depend on chemical properties of the
macromolecule, but only on space dimension $d$ (e.g., the
phenomenological Flory theory \cite{deGennes} gives
$\nu(d)=3/(d+2)$). Thus,
 for polymers in $d=1$-dimensional space,
this exponent takes on the maximal value of $1$ (completely stretched
chain), in $d=2$, one restores the exact value  $3/4$
\cite{Nienhuis82} and in the $3$-dimensional case, the Flory
theory nicely agrees, e.g., with the analytical result $\nu=0.5882 \pm 0.0011$ \cite{Guida}. For space
dimension  $d\geqslant 4$, the polymer behaves as an idealized Gaussian
chain with $\nu=1/2$. The relation of the polymer size exponent
(\ref{RR}) to the correlation length critical index of the
$m$-component spin vector model in the formal limit $m\to 0$ was
provided by  P.-G. de Gennes (the well-known de Gennes limit
\cite{deGennes}).

Note that the Flory theory is applicable only in the case of
polymers in a pure environment, but in reality most of the solutions
contain impurities (obstacles), that interact with the macromolecules.
These obstacles can be very small or penetrate through the whole
space; randomly distributed or obeying some correlations on a
mesoscopic scale \cite{Sahimi95,Dullen79}. It was shown
\cite{Kim87,Kremer81,Lee88} that in the environment having a weak
concentration of quenched point-like obstacles, the macromolecules
behave like in pure solutions, namely the value of the critical exponent $\nu$ in (\ref{RR})
is the same as in the idealized case of a pure solvent. Only when the concentration of defects is close to the
percolation threshold where an
incipient percolation cluster of fractal structure emerges in the system, the scaling properties of polymers are modified in a non-trivial way
\cite{Kremer81,Lee88,Lee89,Chakrabarti81,Sahimi84,Lam84,Lyklema84,Kim87,Cherayil90,Roy90,Lam90,Kim91,Nakanishi91,Vanderzande92,Ordemann00,Janssen07,
Grassberger93,Lee96,Blavatska08_1,Blavatska08_2,Blavatska08_3}.
It is appropriate to mention that according to a simple generalization
of the Flory formula $\nu(d_f)=3/(d_f+2)$ \cite{Kremer81}, there should
be a different behavior for polymers on
 spaces with fractal dimension $d_f<d$, e.g., on percolation clusters.

The conformational properties are also influenced in a non-trivial way
when the position of one obstacle particle affects the other, i.e.,
there are  correlations in the spatial distribution of impurities. In
particular, these correlations often express themselves by a power law
behavior $\sim |r|^{-a}$ \cite{Weinrib83} with $a<d$, where $r$ is the
distance between two obstacles. This type of disorder has a direct
interpretation for integer values of $a$, namely, the cases $a=d-1$
$(a=d-2)$ describe straight lines (planes) of impurities of random
orientation, whereas non-integer values of $a$ are interpreted in
terms of impurities organized in fractal structures. This type of
disorder leads to a new universality class of  polymers
\cite{Blavatska01}.

Herein above we were speaking about polymers in disordered but
isotropic environment. It means that all the observable statistical  properties of the molecules
are the same when explored in different spatial directions. An  interesting question concerns a
quantitative change of these properties when there are
 some selected directions in the system: the case of spatial
anisotropy. In this concern, a widely discussed  model of directed
self-avoiding walks (DSAW) \cite{vanderzande} may describe the
properties of macromolecules in an applied external  field.  This causes
an elongation of the molecule along the field direction
\cite{Bhattacharjee,Baram85} and leads to the existence of two characteristic length scales
 in parallel and perpendicular directions. The
anisotropic properties of the environment can  also be caused by the presence
of extended obstacles  correlated in $\varepsilon_d$-dimensions, similar to
those discussed previously for the spin systems. These may be ordered colloid particles, gel fibers or
biological species in a cell environment. In these systems we
should also expect a different behavior along the chosen direction and
perpendicular to it. Baumgartner and Muthukumar \cite{Baumgaertner96}
discussed a model of polymer chains in an environment having impurities in
the form of absorbing parallel cylinders. They predicted that the
polymer chain is elongated  in the direction parallel to the  cylinders and,
correspondingly, the parallel component of the end-to-end distance is
governed by an exponent $\nu_{||}=1$, whereas in normal direction there
will be no dependence of the end-to-end distance component on the  number
of monomers at all (exponent $\nu_{\perp}=0$). This  gives a
reason to expect an anisotropic behavior for  polymers in solution
having impurities correlated in $\varepsilon_d$ dimensions
\cite{Haydukivska13}.

In the present paper, we study the conformational properties of  polymers in the environment where an anisotropy is caused
by the presence of  $\varepsilon_d$-dimensional defects of parallel orientation, which are randomly distributed in the remaining $d-\varepsilon_d$
dimensions, both using numerical simulations based on a discrete lattice model (Section II) and
 an analytical description within des Cloizeaux direct polymer
renormalization scheme (Section III). We end up by giving conclusions and an outlook.

\section{Numerical studies}

\subsection{The model}

Our goal is to investigate the conformational properties of polymers
in anisotropic environments in the presence of obstacles that are ordered
in some subspace. For this purpose, we choose a lattice model of a
long flexible polymer chain~--- the model of self-avoiding random
walks (SAW), which is established to perfectly capture the universal
properties of polymers in good solvent with the excluded volume effect.
We deal with the cubic lattice since it is known that universal
properties do not depend on the lattice type \cite{desCloizeaux} and
the cubic one is the most simple and easy to realize.

\begin{figure}[!t]
\begin{minipage}[b]{0.5\textwidth}
{\includegraphics[width=\textwidth]{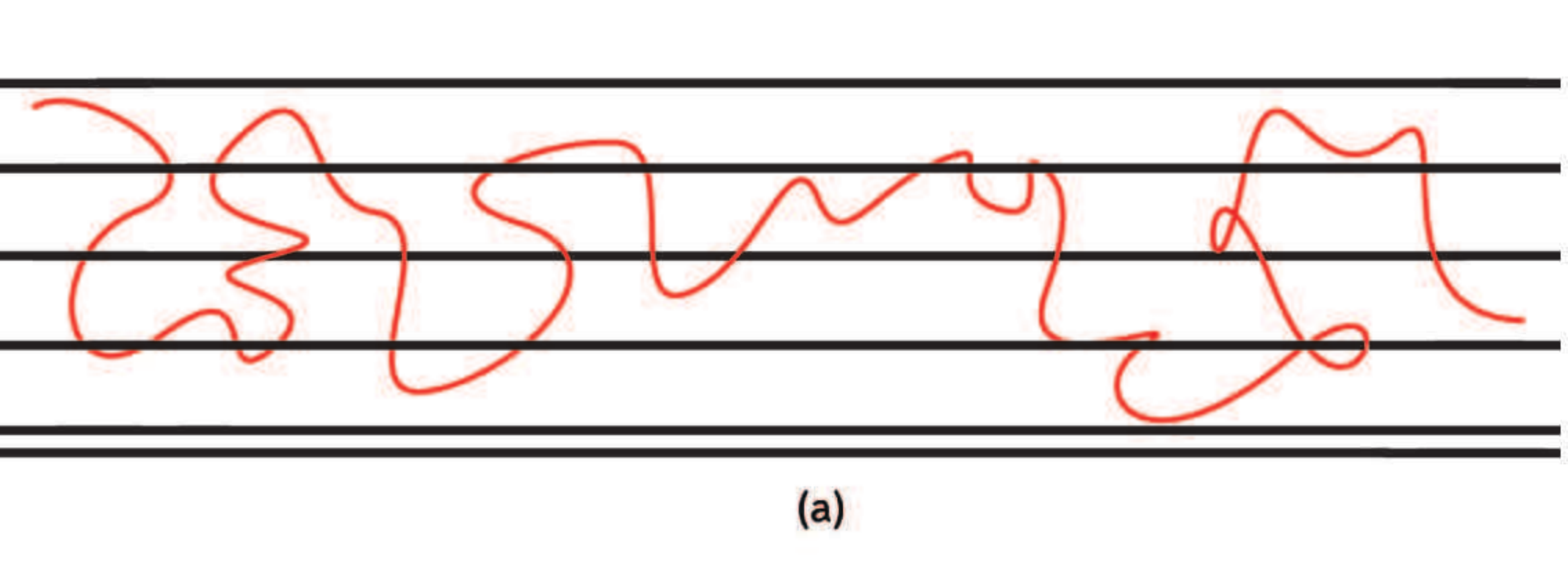}\\[3ex]  
\includegraphics[width=\textwidth]{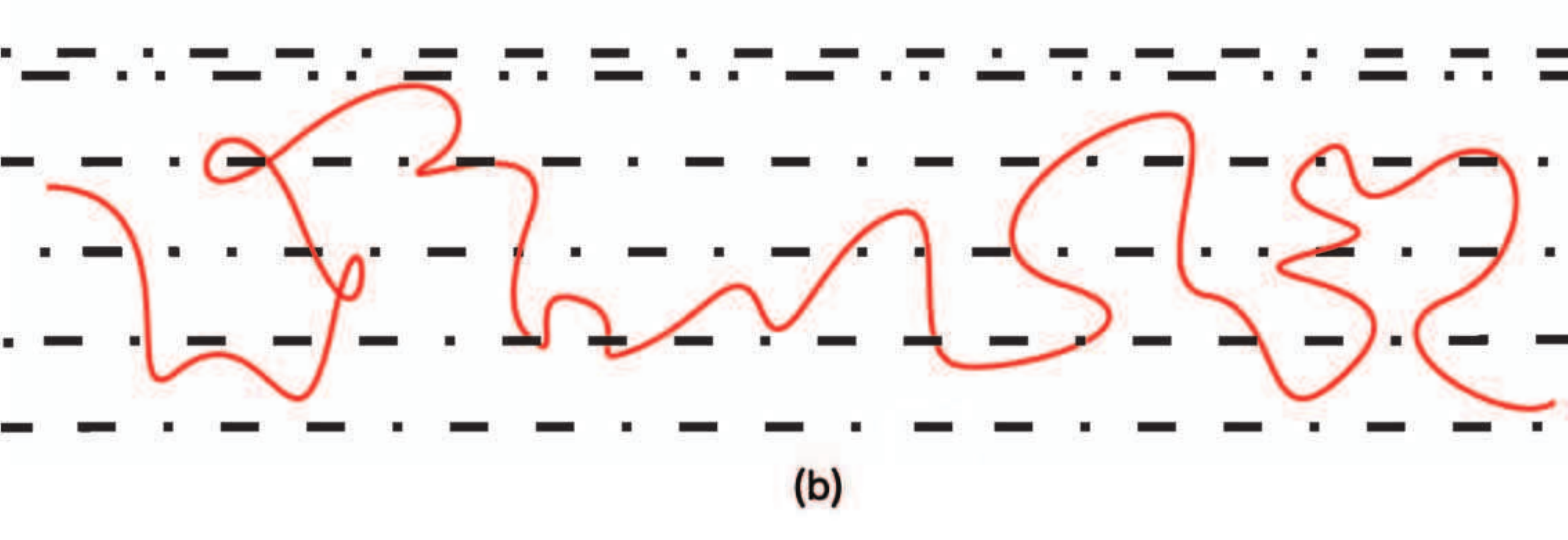}}
\end{minipage}
\hfill
\begin{minipage}[b]{0.5\textwidth}
\includegraphics[width=\textwidth]{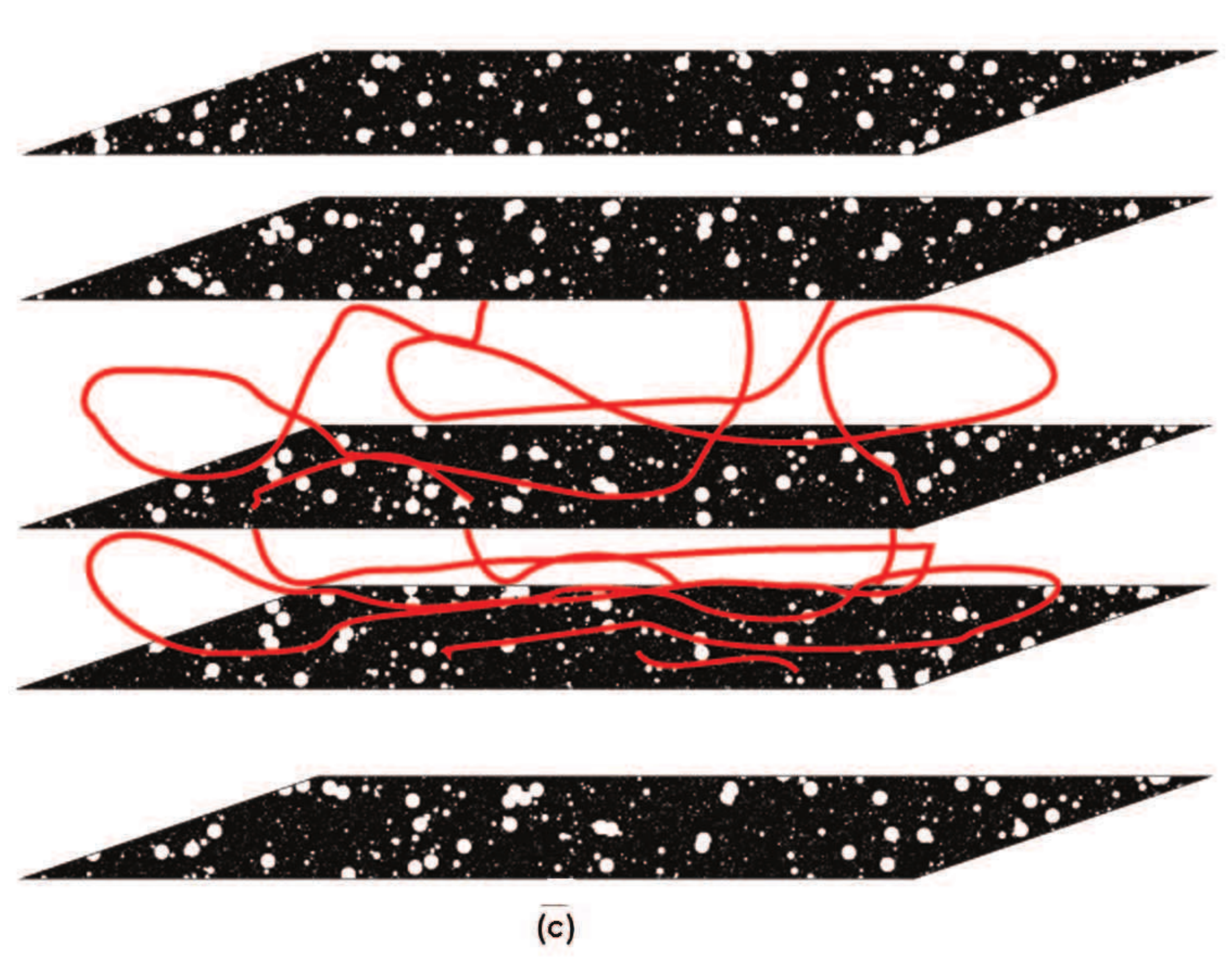}
\end{minipage}
\caption{(Color online) Schematic representation of polymer chain in the environment having structural obstacles in the form of lines (a), partially penetrable lines (b) and
 partially penetrable planes (c).} \label{fig:0}
\end{figure}

The simplest types of extended obstacles that can be chosen for our purposes are spacial
objects in the form of lines ($\varepsilon_d=1$) of parallel orientation (see figure~\ref{fig:0}~(a)),
spreading throughout the lattice in some chosen direction,
since they should obviously lead to a different behavior in directions parallel
and perpendicular to them. The case of homogeneous planes is not of interest since
they divide the space into small restricted regions, and thus the problem is reduced to that
of polymers in confined geometries.

\begin{figure}[!b]
\begin{center}
\includegraphics[width=90mm]{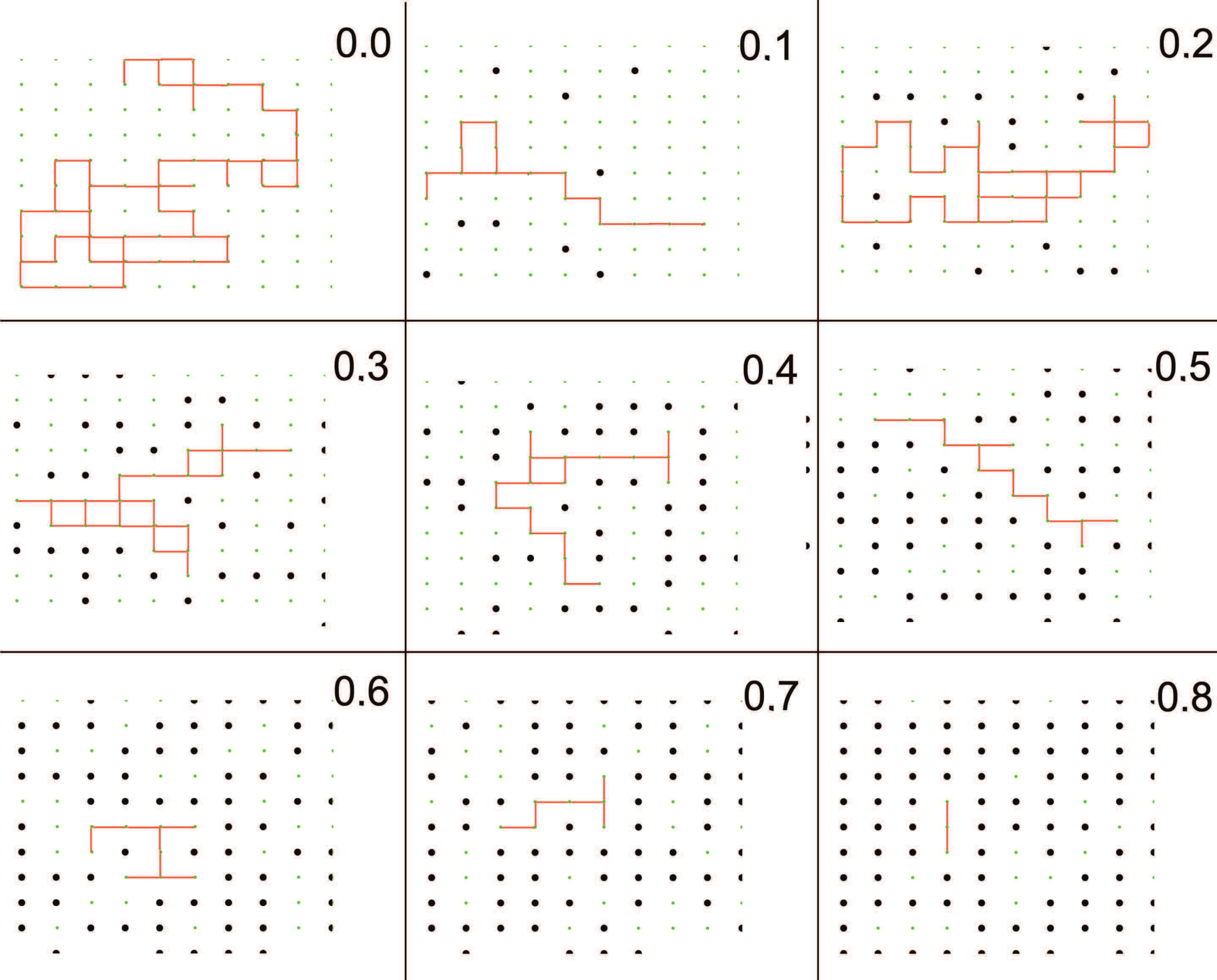}
\end
{center}
\caption{(Color online) Projections of SAW trajectories on the $xy$-plane of a
lattice at different concentrations $c$ of impurity lines extended
throughout the system in $z$ direction. } \label{mess}
\end{figure}

We start with an $xy$-plane of a lattice  with concentration $c$ of
randomly chosen sites containing point-like defects and then we
build lines perpendicular to this plane in $z$-direction through
chosen sites. In figure~\ref{mess} one can see a projection of the SAW
trajectory on this plane for different concentrations of lines. As
one can see at high concentrations $c$ of such obstacles, the SAW
trajectory appears to collapse in small restricted regions. In this case, a
long polymer chain will behave like a one-dimensional rod, extended in
the direction parallel to the lines of defects. Thus, we expect  a
crossover to an extended regime  for an increasing   defect
concentration.

We also consider more interesting situations, namely a set of
partially penetrable lines (figure~\ref{fig:0}~(b)) and planes (figure~\ref{fig:0}~(c)). To this end, with some fixed probability  $p$, we randomly choose the sites on the constructed lines (planes)
of defects and treat them as ``open'' (allowed for SAW trajectory).
Note that one can, roughly speaking, treat these objects as fractals.
For example, the fractal (cluster) dimension of such a line with concentration $p$
of the lacking sites can
be estimated from the well-known relation between
the linear size of an object and ``the number of particles'':
$\varepsilon_d\cong\ln[(1-p)\cdot L]/\ln(L)$ (here, $L$ is the length of a line).  Indeed, at $p=0$, one
has a Euclidian line (like in figure~\ref{fig:0}~(a)) and simply restores $\varepsilon_d=1$,
 whereas with increasing $p$, the line can be treated as a set of disconnected sites (points) and the fractal dimension of this
  object gradually tends to 0.

\subsection{The method}

For our purposes we use the Pruned-Enriched Rosenbluth Method
(PERM) \cite{Grassberger97}. It is based on the original
Rosenbluth-Rosenbluth algorithm of growing chains with population
control parameters \cite{Rosenbluth55}.
On  each step $n$, the chain has a weight $W_n$ given by:
\begin{equation}
W_n=\prod_{i=1}^n m_i\,, \label{weight}
\end{equation}
where $m_i$ is the number of possibilities to perform the next step,
which varies from $0$ to $2d-1$ due to the fact that the chain may not
cross itself. This value is also reduced by the presence of impurities.

When the chain of total length $N$  is constructed, a new one is started from the same starting point, until the desired number of chain
configurations are obtained.
In this manner, all observables should be averaged over an ensemble of different
chain configurations $M$:
\begin{equation}
\langle (\ldots)
\rangle=\frac{1}{Z_N}{\sum_{k=1}^{M}W_N^{k}(\ldots)}\,,
\qquad Z_N=\sum_{k=1}^{M} W_N^{k}\,,
\end{equation}
here, $W_N^{k}$ is the weight of the $k$-th configuration of the $N$-step trajectory.

It is also  necessary to average over different configurations of
disorder:
\begin{equation}
\overline {(\ldots)}=\frac{1}{p}{\sum_{k=1}^{p}(\ldots)},\label{average}
\end{equation}
where $p$ is the number of replicas (the number of different realizations of the disorder).
For our purposes, we consider about $10^5$ chains for each of the $400$
replicas.

Weight fluctuations are reduced by using population control (pruning
and enrichment). It means that with probability of $1/2$ we reject
the chains having low weight and enrich the statistics by replication increasing the
number of high weighted configurations. To do this, we use  lower
and upper bound  weights, that are updated at each step according
to \cite{Hsu03,Bachmann03_1,Bachmann03_2}: $W_n^{>}=C(Z_n/Z_1)(c_n/c_1)^2$ and
$W_n^{<}=0.2W_n^{>}$, where $c_n$ is the number of  chains of
length $n$ created, and the parameter $C$ controls the pruning-enrichment
statistics; it is chosen in such a way to allow one to receive in
average $10$ chains of total length $N$ in each tour.

\subsection{Results}

We concentrate on the conformational properties of
long flexible polymer chains, in particular, in the critical exponents
governing the behavior of the effective linear size of the macromolecule with
respect to the number of monomers (\ref{RR}).
Let us note that in the anisotropic case we expect to have two different exponents rather than one:
\begin{eqnarray}
&&\overline{\langle R_{||}^2 \rangle} =(z_N-z_0)^2\sim N^{2 \nu_{||}}, \nonumber\\
&&\overline{\langle R_{\bot}^2\rangle} =(x_N-x_0)^2+(y_N-y_0)^2\sim N^{2
\nu_{\bot}},\label{R1}
\end{eqnarray}
so that $R^2=R_{\|}^2+R_{\bot}^2$. The parallel and perpendicular
components  are expected to behave in a different way
\cite{Baumgaertner96}.

\begin{figure}[!t]
\centerline{
\includegraphics[width=0.51\textwidth]{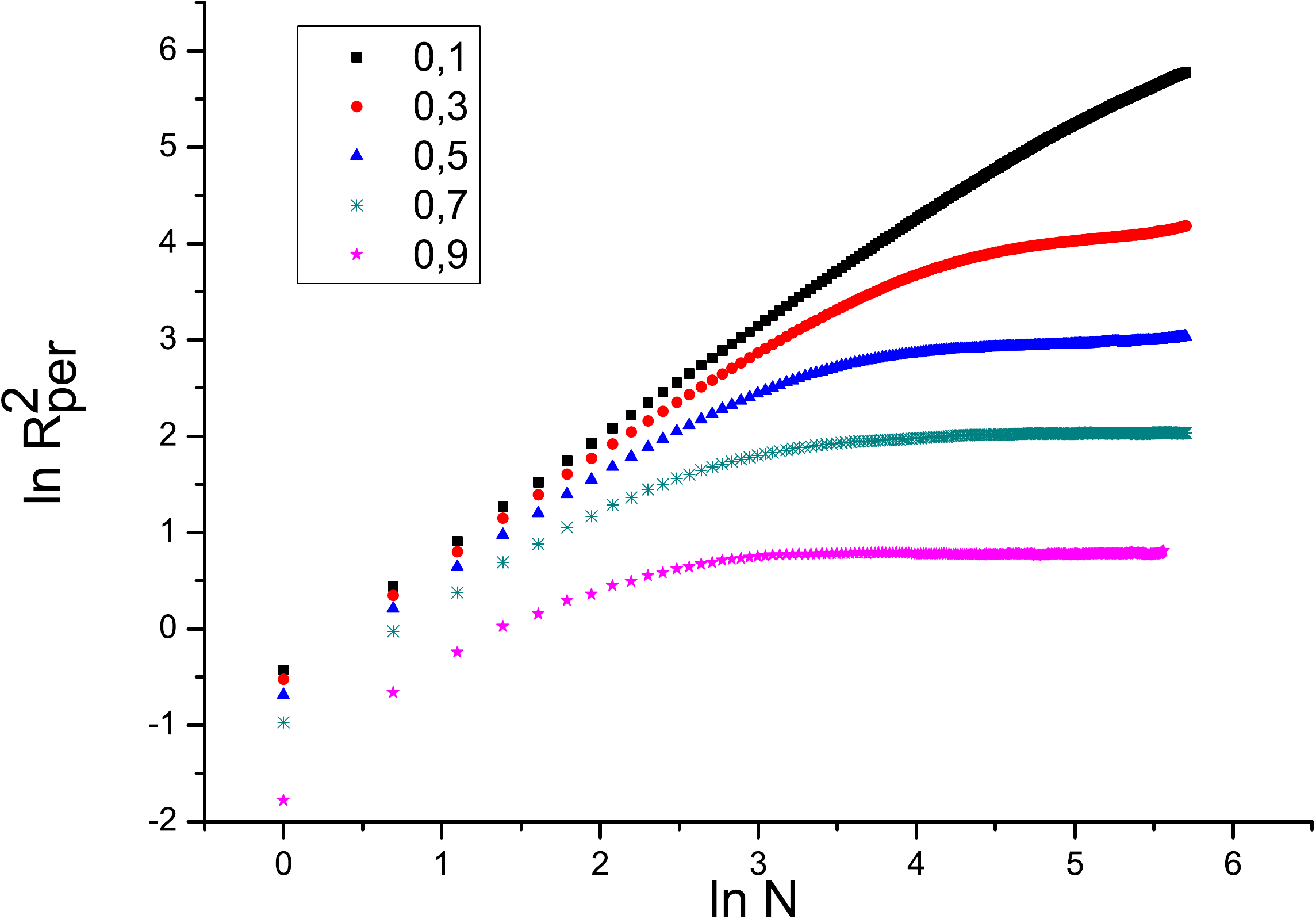}
\hspace{2mm}
\includegraphics[width=0.51\textwidth]{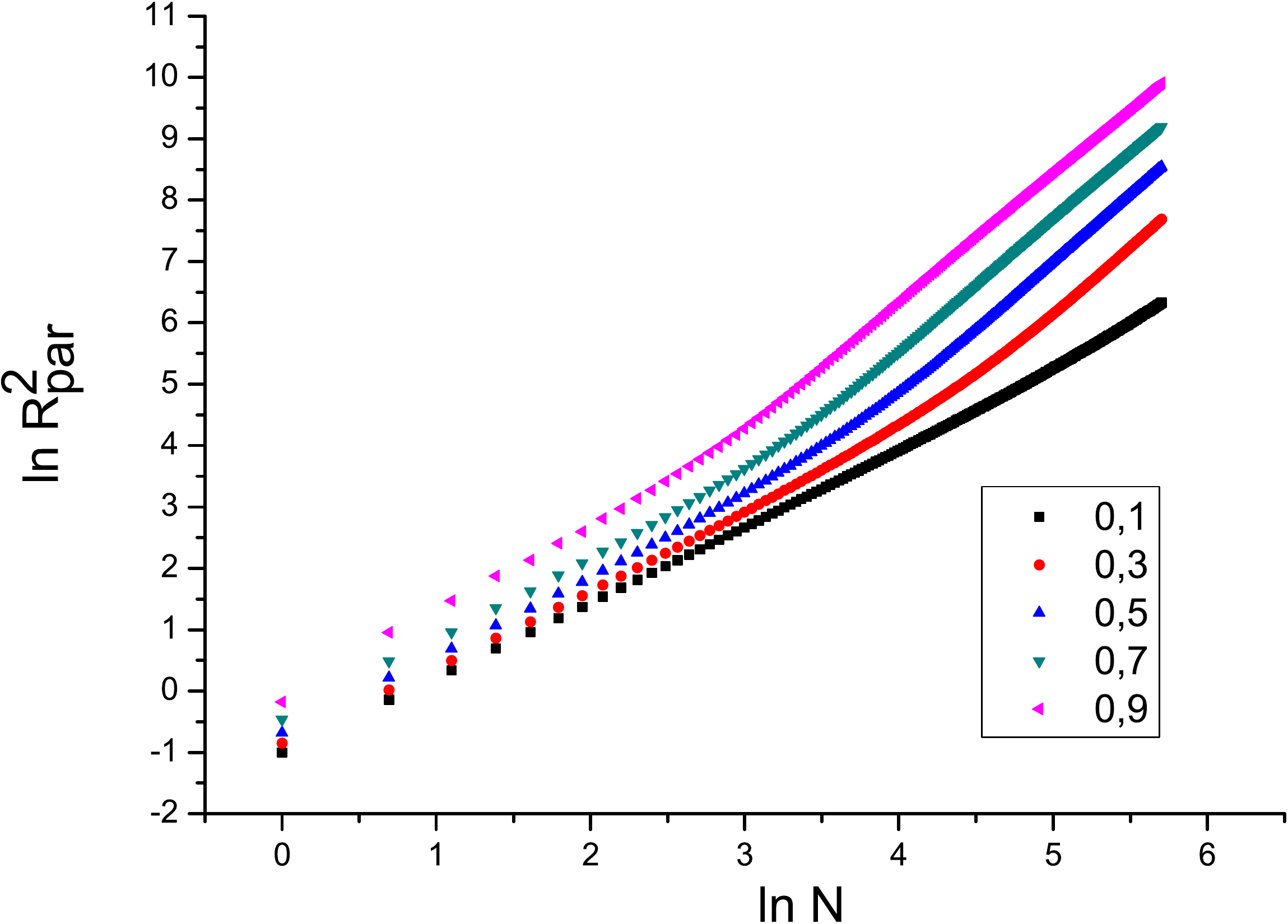}
}
\caption{(Color online) Parallel (a) and perpendicular (b) components of the
end-to-end distance of SAW as a function of chain length in a double logarithmic scale
at various concentrations
of defects in the form of parallel lines. } \label{fig:4}
\end{figure}

\begin{figure}[!b]
\begin{center}
\includegraphics[width=0.5\textwidth]{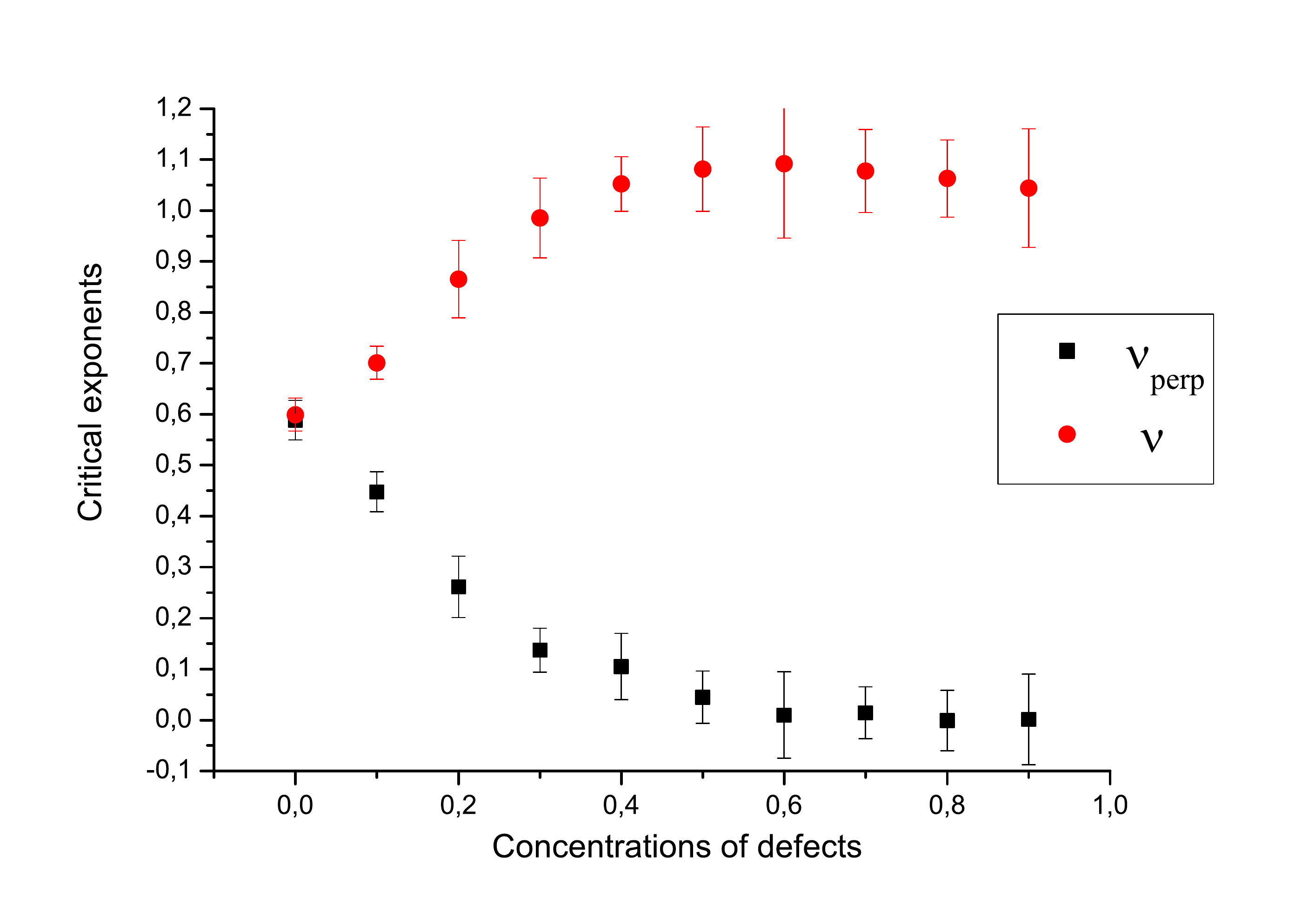}
\end{center}
\caption{(Color online) Critical exponents $\nu_{\|}$ and  $\nu_{\bot}$ governing the
components of the end-to-end distance of SAWs,  parallel and
perpendicular to the defects in the form of parallel lines, as a function of the concentration of defects.} \label{fig:3}
\end{figure}

We start by considering the case of the anisotropy caused by the
presence of obstacles in the form of impenetrable parallel lines
\cite{Haydukivska13}. In this case, we took chains up to $N=300$
monomers and find the components of the end-to-end distance vector,
performing double averaging according to (\ref {average}). The results
are presented in figure~\ref{fig:4}. It is clearly observed that the
behavior of the two components differ from each other: the parallel component grows with an increase of concentration of obstacles, whereas  the perpendicular component collapses indicating the stretching of the polymer chain
in longitudinal direction.  As one can see,
there is a crossover between the two types of behavior: one for short
chains (less than $40$ monomers) and then the other one for long
polymers. We expect this crossover to take place when the averaged
end-to-end distance $R^2$ of the SAW trajectory  is comparable to the
distance between the lines of impurities. This is similar to the polymer
behavior in a restricted space, for example a cylinder, when short
chains (with end-to-end distance smaller than the  radius of
cylinder) behave like $3$-dimensional, and long chains as
$1$-dimensional \cite{deGennes}. In our case, we observe a crossover
to such a behavior when the concentration of lines that penetrate
the system is close to a percolative concentration of
point-like defects on a simple square lattice. At smaller
concentrations, applying the least-square fitting of the data observed,
we obtain the estimates for the two critical exponents $\nu_{\|}$ and
$\nu_{\bot}$  (see figure~\ref{fig:3}), which coincide  at $c=0$, where
we restore the corresponding value
 of the SAW exponent on a pure lattice.
As one can see, the exponent governing the scaling of a parallel component
of the end-to-end distance is larger than the pure one and gradually reaches the maximal value of 1 with an increasing
concentration of disorder, while $\nu_{\bot}$  is smaller and gradually tends to zero. It gives us the right to say that polymers
in anisotropic space are more elongated than  polymers in isotropic
environments.

Let us check the fact of possible elongation by analyzing the shape
properties. All information concerning the shape measure of a chain is given by the
gyration tensor with its components defined by:
\begin{equation}
Q_{\alpha\beta}=\frac{1}{N^2}\sum_{i=1}^N\sum_{i=1}^N
(x^{\alpha}_i-x^{\alpha}_j)(x^{\beta}_i-x^{\beta}_j),\qquad \alpha,\beta=1,\ldots,3,
\end{equation}
where $x^{\alpha},x^{\beta}$ are the components of the position vector
$\vec{r_{i}}$  of the $i$-th monomer. Eigenvalues
of this tensor  provide a full information on the shape of the
polymer. To receive these in simulations we need to solve a cubic
equation for each of the $10^5$ chains. Thus, we are interested in
calculating  the rotationally invariant shape characteristics, such as
asphericity and prolateness, defined in $d=3$ as combinations of
gyration tensor components according to \cite{Solc71_1,Solc71_2,Aronovitz86}:
\begin{equation}
A_3=\frac{3}{2}\frac{{\textrm{Tr} }\,\hat{Q}^{2}}{({\textrm{Tr}}\, Q)^{2}}\,,\qquad
S_3=27\frac{{\textrm{det}}\, \hat{Q}^{2}}{({\textrm{Tr}} Q)^{2}}\,,
\end{equation}
with $\hat{Q}=Q-\hat{I}{{\textrm{Tr} Q}}/{3}$ ($\hat{I}$ being a unity matrix).
Asphericity is normalized in such a way that it changes the value from $0$ for spherical configuration
to $1$ for a completely stretched rod-like structure. Prolateness obeys the  inequality: $-0.25\leqslant S \leqslant 2$,
it is negative for oblate configurations and positive for
prolate ones;  a value $2$ corresponds to a rod-like (completely prolate) state.

\begin{figure}[!b]
\begin{center}
\includegraphics[width=0.49\textwidth]{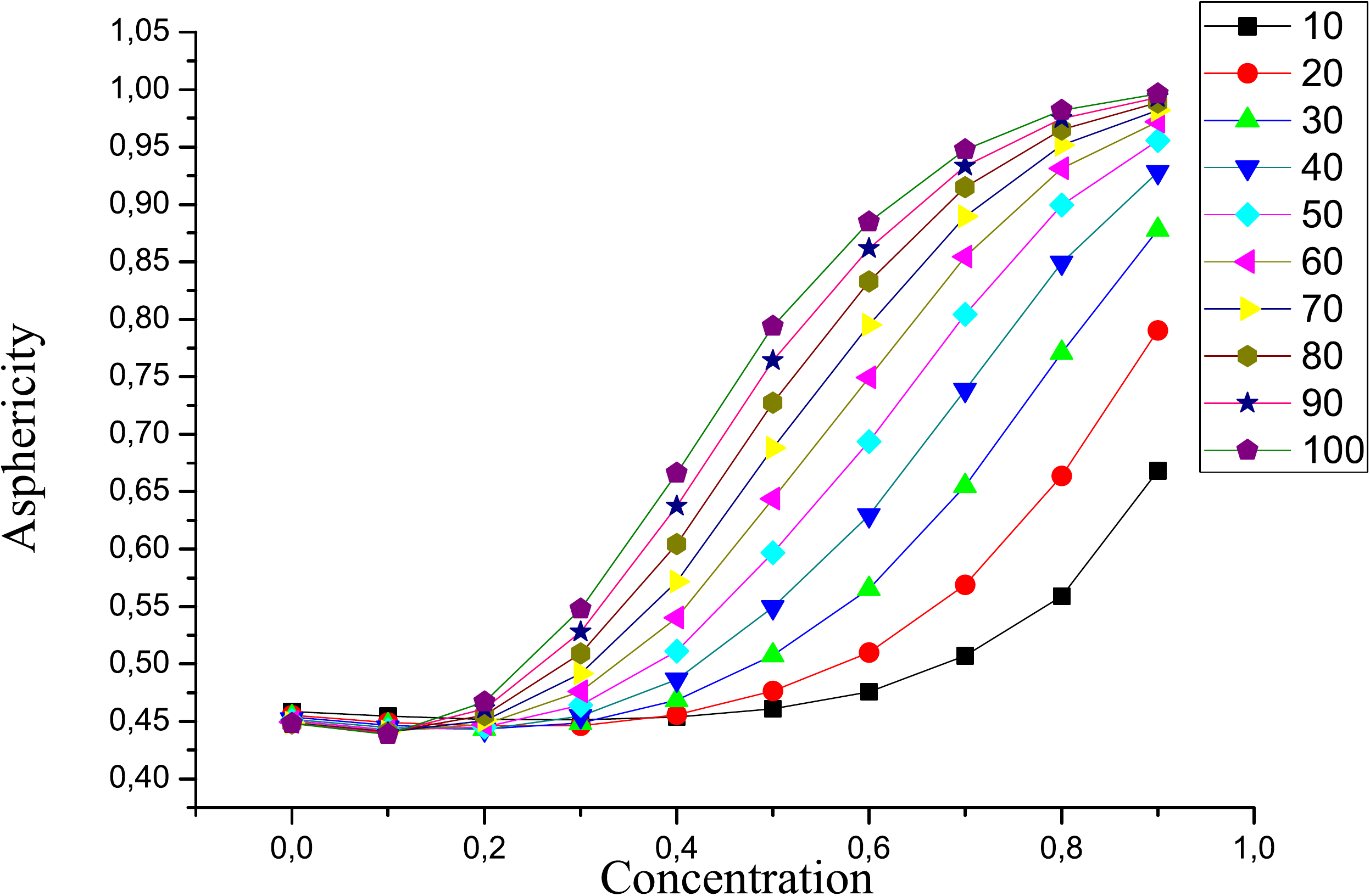}
\hfill
\includegraphics[width=0.49\textwidth]{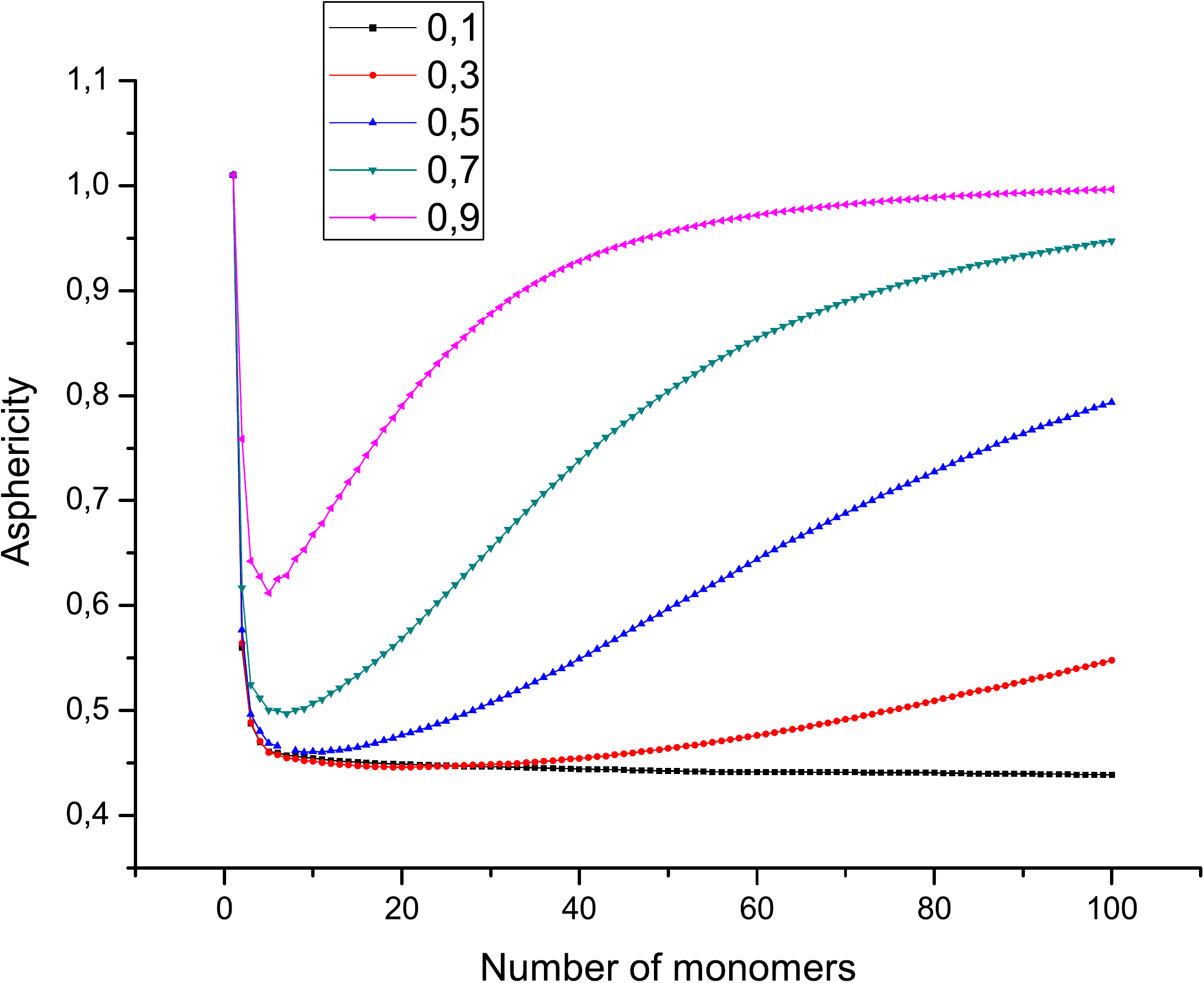}
\end{center}
\caption{(Color online) Asphericity of SAW trajectories as a function of the concentration of defects at various numbers of monomers (a). Asphericity  of SAW
trajectories as a function of the number of monomers at various fixed concentrations of defects (b).} \label{fig:9}
\end{figure}

\begin{figure}[!t]
\begin{center}
\includegraphics[width=0.49\textwidth]{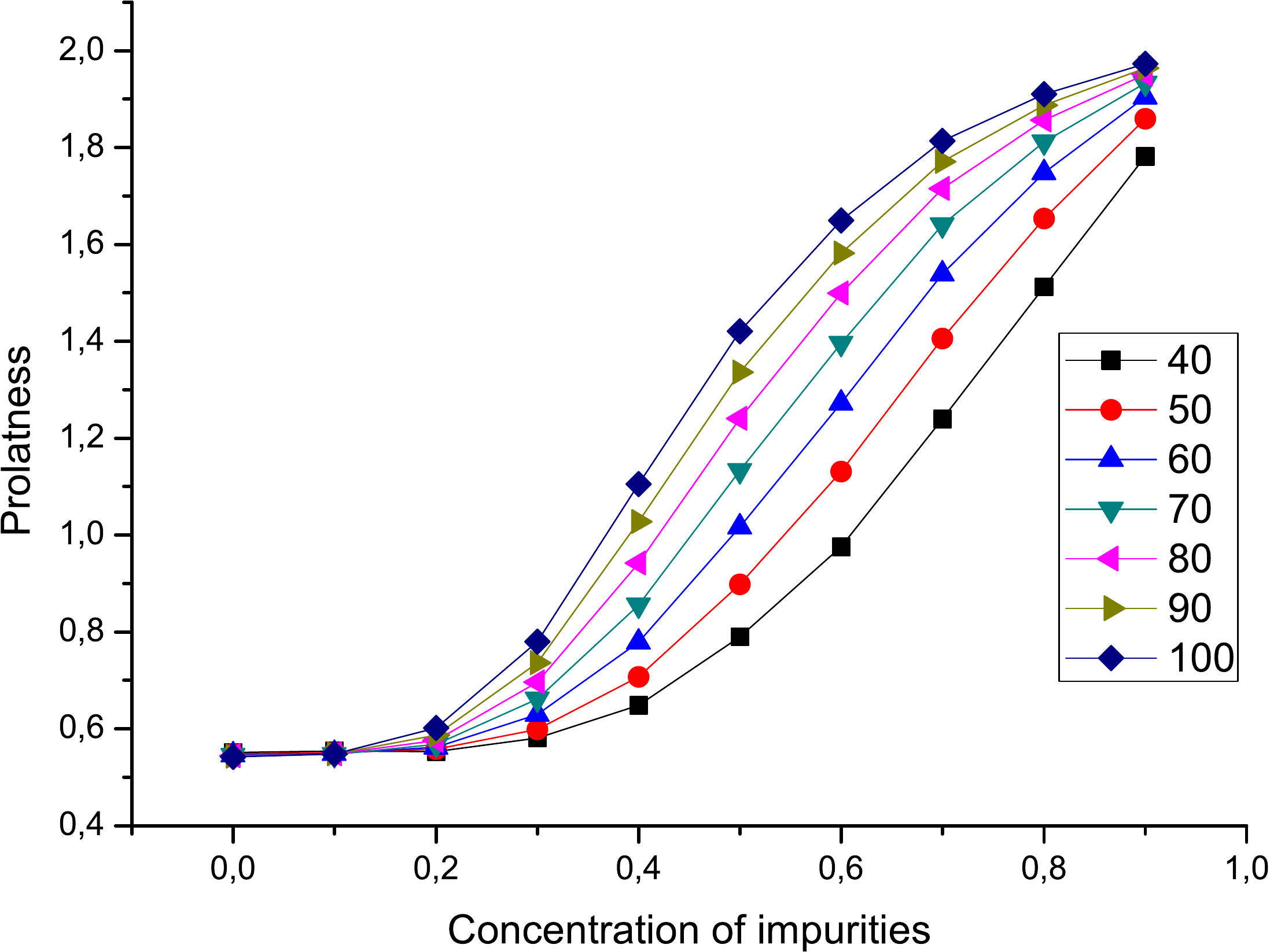}
\hfill
\includegraphics[width=0.49\textwidth]{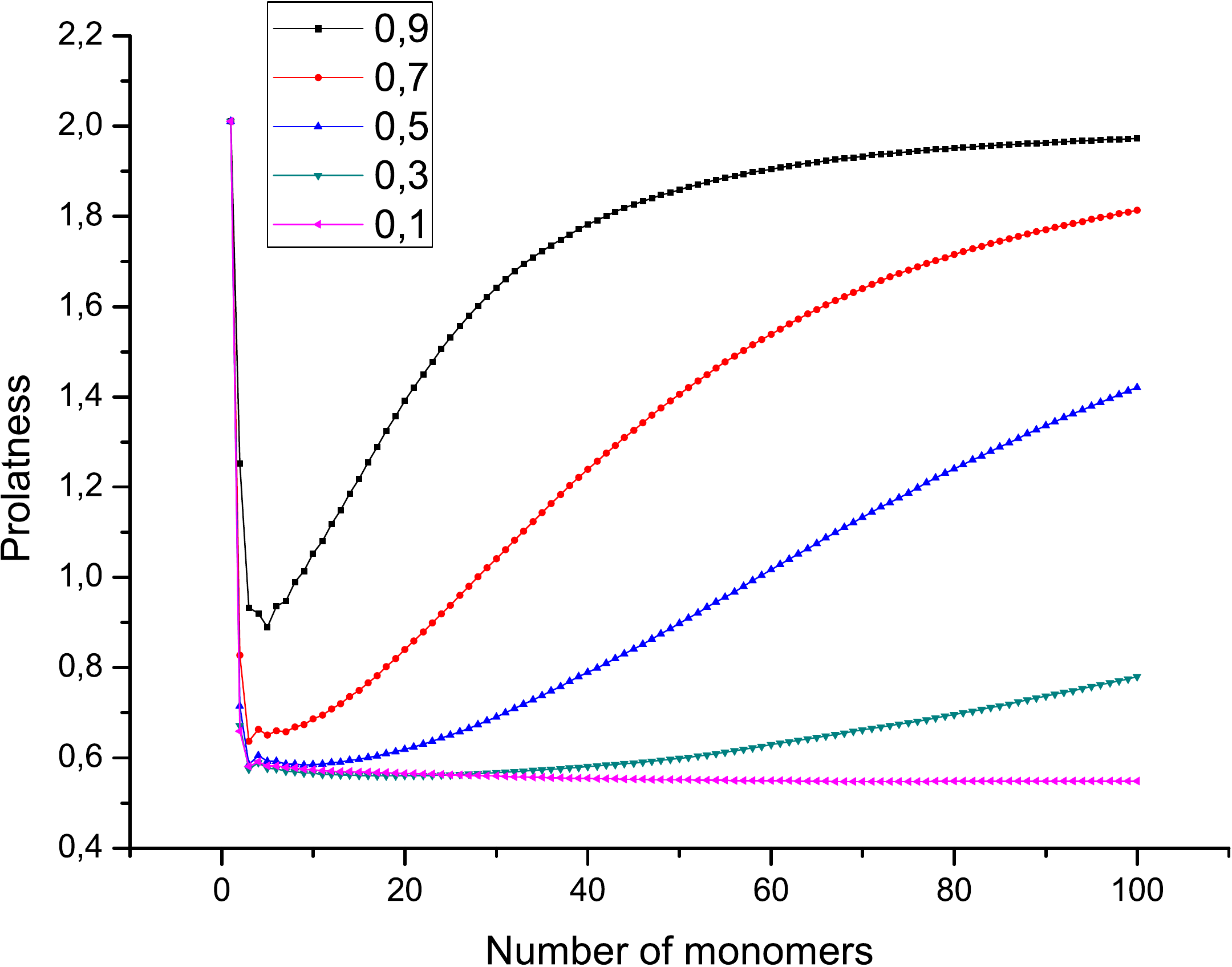}
\end{center}
\caption{(Color online) Prolateness of SAW trajectories as a function of the concentration
of defects at various numbers of monomers (a). Prolateness  of SAW
trajectories as a function of the number of monomers at various fixed
concentrations of defects (b).} \label{fig:7}
\end{figure}

In figures~\ref{fig:9} and \ref{fig:7} we present our data for
$\overline{\langle A_3 \rangle}$ and $\overline{\langle S_3
\rangle}$, averaged over realizations of disorder at various fixed
concentrations of defects. At $c=0$, in both cases we restore the
corresponding values on a pure lattice. One can easily see in
figures~\ref{fig:9}~(a) and \ref{fig:7}~(a) that both quantities are growing with
an increasing defect concentration  and gradually reach the
corresponding values of rod-like structures.

Next, we consider an interesting situation, where point-like defects
in the lattice are aligned in some particular direction (say, $z$)
forming partially penetrable lines (see figure~\ref{fig:0}~(b)). To
investigate the influence of  such a type of space anisotropy on the
conformational properties of flexible polymer chains, we randomly
choose some concentration $p$ of sites on lines of defects
(constructed as described previously), and treat them as
``open'' (allowed to SAW trajectory). As a result, we obtain aligned
fractal-like objects with dimension $0<\varepsilon_d<1$. In such problems, we have two
parameters: the concentration $c$ of obstacles in the form of
parallel lines, and  the probability $p$ of a SAW trajectory to
penetrate through this line.

\begin{figure}[!b]
\begin{center}
\includegraphics[width=0.49\textwidth]{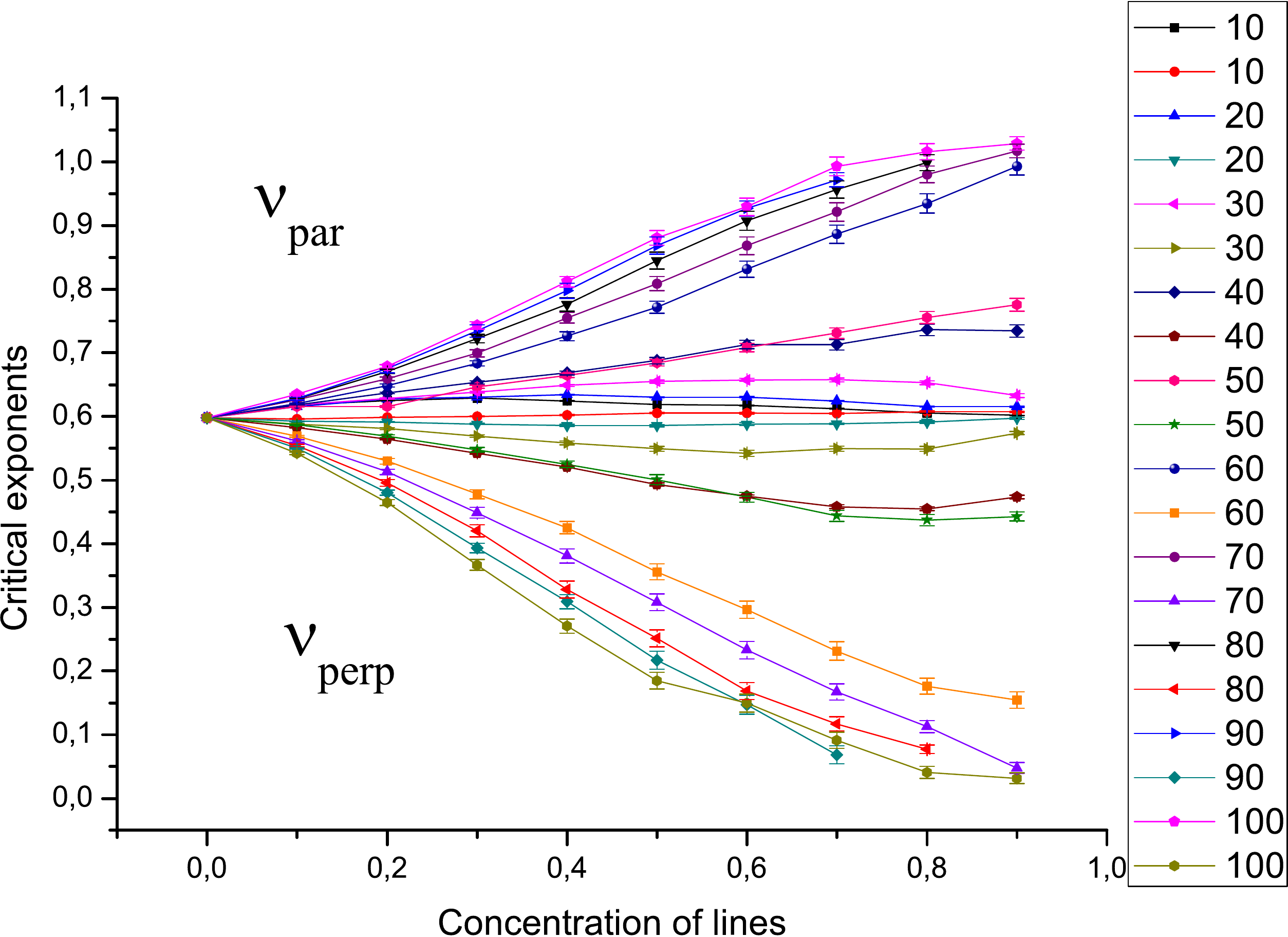}
\hspace{1mm}
\includegraphics[width=0.49\textwidth]{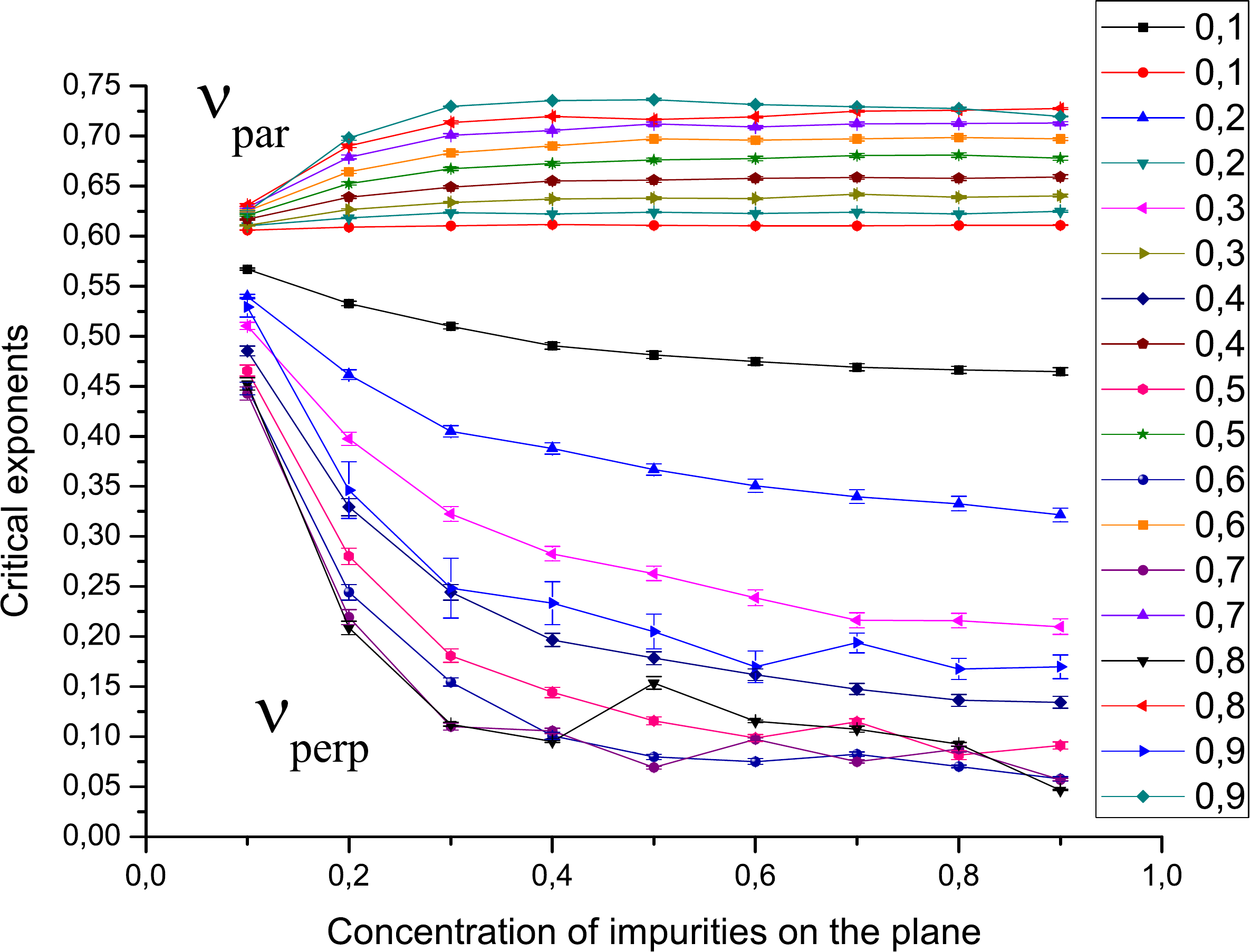}
\end{center}
\caption{(Color online) Critical exponents $\nu_{\|}$ and $\nu_{\bot}$ of SAW on a
lattice with partially penetrable lines of
defects as a function of the concentration of lines at various probabilities to penetrate through these lines (a). Critical exponents $\nu_{\|}$ and $\nu_{\bot}$ of SAW on a
lattice with partially penetrable planes of
defects  as a function of  probabilities to penetrate these lines at various concentrations of planes (b).} \label{fig:11}
\end{figure}

Performing simulations for chains up to
$N=100$ steps and applying the least-square fits for the data
obtained for parallel and perpendicular components of the end-to-end
distances of polymer chains, we received the estimates for critical
exponents $\nu_{\|}$ and  $\nu_{\bot}$  (see figure~\ref{fig:11}~(a)) as
functions of these two parameters. Again, the exponent governing the
scaling of  a parallel component of the end-to-end distance is larger
than the pure one and gradually reaches the maximal value of 1, while
the other one is lower and gradually tends to zero.  This tendency
is kept even at a rather high
probability of a growing trajectory to penetrate the line (up to the concentration $p$ of ``open'' sites close
to a critical percolation concentration).
When $p$ is close to 1, the spatial anisotropy disappears and both exponents gradually
reach the corresponding value of the  pure lattice case.

Finally, we consider another possible model of  anisotropic
environment caused by the presence of structural defects in  the form of
partially penetrable planes of parallel orientation (see figure~\ref{fig:0}~(c)). We start with homogeneous planes of concentration
$c$, randomly distributed in $z$-direction, and again randomly
choose some concentration $p$ of sites on these planes and treat
them as ``open''(allowed to SAW trajectory). As a result, we obtain
fractal-like objects having dimension $1<\varepsilon_d<2$. Performing simulations for
chains up to $N=100$ steps and applying the least-square fits for
data obtained for parallel and perpendicular components of the
end-to-end distances of polymer chains, we receive  estimates for the
critical exponents $\nu_{\|}$ and  $\nu_{\bot}$  (see figure~\ref{fig:11}~(b)) as functions of these two parameters.
The exponent governing the scaling of  the parallel component of the
end-to-end distance gradually changes from the value on
three-dimensional pure lattice (at $c=0$) to that in two dimensions
with a growing concentration of impurity planes. This can be treated
as a crossover to a restricted geometry regime of polymers confined
between two planes. The exponent $\nu_{\bot}$ gradually tends to
zero.

\section{Analytical approach}

\subsection{The model}

We deal with  flexible polymers in an environment with extended
impurities correlated in $\varepsilon_d$-dimen\-sions and randomly
distributed in the remaining space. Let us start with a continuous
x`model, where a polymer chain is presented as a path of  length (or surface) $S$
parameterized by $\vec{r}(s)$, with $s=0\ldots S$ \cite{Edwards}.
An effective Hamiltonian of the system is given by:
\begin{eqnarray}
&&H=\frac{1}{2}
\int^{S}_{0}\left(\frac{\rd\vec{r}(s)}{\rd s}\right)^{2}\,\rd s
+u\int^{S}_{0}\rd s'\int^{s'}_{0}\rd s''\delta\left(\vec{r}(s')-\vec{r}(s'')\right)\,\rd s
+\int^{S}_{0}V(\vec{r}(s))\,\rd s .
\end{eqnarray}
Here, the first term describes the chain connectivity, the second
term reflects the short-range repulsion between monomers due to the
excluded volume effect with coupling constant $u$,
and the last term arises due to the interaction between the monomers of the
polymer chain and the structural defects in the environment given by
potential $V$. We work in the formalism of partition functions:
\[
Z=\int \textrm{D}\vec{r} \re^{-H},
\]
where $\int \textrm{D}\vec{r}$ denotes integration over different paths.

Dealing with systems that display randomness of structure, one
usually encounters two types of ensemble averaging, treated as
annealed and quenched disorder \cite{Brout59_1,Brout59_2}. In general, the
critical behavior of disordered systems with annealed and quenched
averaging is quite different. However, for the polymer systems it
has been shown \cite{Cherayil90,Wu91,Ippolito98,Patel03,Blavatska13}, that the
distinction between quenched and annealed averages for an infinitely
long single polymer chain  is negligible, and in performing
analytical calculations for quenched polymer systems one may
restrict the problem  to the simpler case of annealed averaging. In
this paper, we deal with annealed averaging over disorder because it is technically simpler. After averaging the
partition sum over realizations of disorder
 and including only up to the second moment of cumulant
 expansion, we receive:
\begin{eqnarray}
\overline{\exp\left\{\int_0^S V(\vec{r}(s'))\rd s'\right\}}&=&
 \exp\left\{\int_0^S \overline{V(\vec{r}(s'))}\rd s'\right\} \nonumber\\
&& \times \exp\left\{\frac{1}{2}\int_0^S \int_0^{s'}
\overline{V(\vec{r}(s'))V(\vec{r}(s''))}-\overline{V(\vec{r}(s'))}^2\rd s'\rd s''\right\},\nonumber
\end{eqnarray}
where $\overline{V(\vec{r}(s'))}$ gives the average concentration of
impurities $\rho$, and for the second moment we assume
\cite{Baumgaertner96,Dorogovtsev80}:
\begin{equation}
\overline{V(\vec{r}(s'))
V(\vec{r}(s''))}=v\,\delta^{d-\varepsilon_d}\left(r_{d-\varepsilon_d}(s')-r_{d-\varepsilon_d}(s'')\right),
\end{equation}
which reflects the fact that the
impurities are correlated in $\varepsilon_d$ dimensions and
uncorrelated in the remaining space.
Omitting the terms $\rho S+\frac{1}{2}\rho^2S^2$, which
give a trivial constant shift, we obtain an averaged partition function $ {\overline Z} =\int d\vec{r} \re^{-H^\textrm{eff}}
$
 with an effective Hamiltonian:
\begin{eqnarray}
H^\textrm{eff}&=&\frac{1}{2}\int^{S}_{0}\left(\frac{\rd\vec{r}(S)}{\rd s}\right)^{2}\rd s +u\int^{S}_{0}\rd s'\int^{s'}_{0}\rd s''
\delta\left(\vec{r}(s')-\vec{r}(s'')\right)\rd s \nonumber\\
&&-v\int^{S}_{0}\rd s'\int^{s'}_{0}\rd s''\delta^{d-\varepsilon_d}\left(r_{d-\varepsilon_d}(s')-r_{d-\varepsilon_d}(s'')\right).
\label{H}
\end{eqnarray}
Note that the last term in (\ref{H}) describes an effective attractive interaction
between monomers in the direction perpendicular to the extended obstacles governed by a coupling constant $v$.

\subsection{The method}

Within the model, all the parameters depend on the polymer area $S$ and
on dimensionless coupling constants $\{z_a\}=a
(2\pi)^{-(d_a)/2}S^{4-(d_a)/2}$ (here $d_a$ is dimension of
coupling constant $a$) in a way that when $\{z_a\}=0$, one restores the case of idealized  Gaussian chain
without any interactions between monomers. To calculate the universal properties of the model we need to
find such values of parameters that lead to physical values of the
universal characteristics. For that reason, we use the direct
renormalization technique proposed by des Cloiseaux
\cite{desCloizeaux}.

Within this scheme, renormalized coupling constants are defined by:
\begin{equation}
g_a(\{z_a\})=-[\chi_1(\{z_a\})]^{-4}Z_{z_a}(S,S)[2 \pi
\chi_0\{z_a\}S]^{-(2-\varepsilon_a/2)}\,, \label{cc}
\end{equation}
where $\varepsilon_a$ is the deviation from the upper space dimension for the
coupling constant $z_a$. $\chi_0=R^2/Sd$ is the swelling factor that
governs the behavior of the end-to-end distance of the polymer in
solution. It can be presented as a perturbation theory series
over the coupling constants:
\begin{equation}
\chi_0(\{z_a\})=1+\sum_a z_a\cdot f_a(d_a),
\end{equation}
where the first term ($1$) corresponds to an ideal Gaussian chain and the others
give corrections caused by interactions in the system.
$f_a(d_a)$ is the factor that depends only on the dimension of the
corresponding coupling constant. This factor allows us to estimate
the critical exponent $\nu$ using the relation:
\[
2\nu-1=S\frac{\partial}{\partial S} \chi_0(\{z_a\}).
\]
The factor $\chi_1$ is connected with the  partition function
$Z(S)/Z^0(S)=[\chi_1(\{z_a\})]^2$. Here, $Z^0(S)$ is the partition
function of an idealized Gaussian chain,   $Z_{z_a}(S,S)$ is the
partition function of two interacting polymer chains.

Renormalized coupling constants given by the equation (\ref{cc})
tend to constant values or the so-called fixed points as the polymer area
tends to infinity. The fixed points are defined as
 common zeros of the flow equations:
\begin{equation}
W_a=2 S \frac{\partial}{\partial S}z^{\ast}_a(\{z_a\}).\label{w}
\end{equation}
To find the fixed points of the model, one need to express $z_a$ in terms
of $g_a$ and find the common zeros of (\ref{w}) and then choose those  that are stable and physical in the region
of interest for the parameters
$\varepsilon_a$.

\subsection{Results}

\begin{figure}[!t]
\begin{center}
\includegraphics[width=100mm]{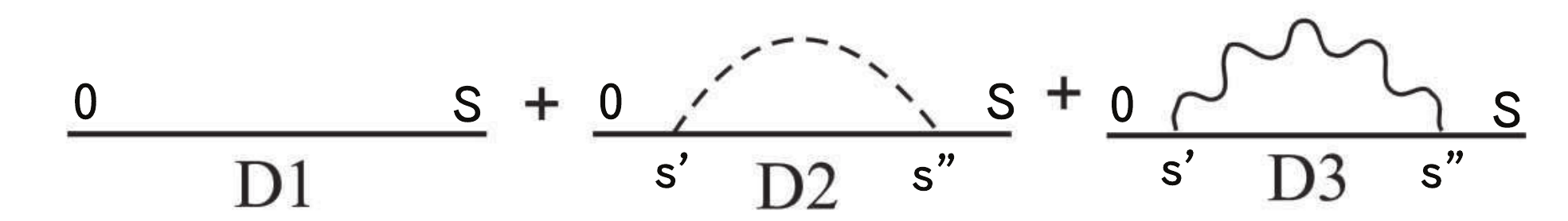}
\end{center}
\caption{Diagrammatic presentation of the contributions to the restricted
partition function (\ref{Zk}) up to the  first order of perturbation theory expansion in the
coupling constants.} \label{fig:1}
\end{figure}

We start with the restricted partition function
\begin{equation}
\widetilde{Z}(\vec{k},S)={\overline {\left\langle {\rm e}^{\ri\vec{k}(\vec{r}(S)-\vec{r}(0)}\right\rangle}}\, ,
\label{Zk}
\end{equation}
where ${\overline {\langle (\ldots) \rangle}}$ means averaging with the  hamiltonian (\ref{H}).
We consider the evaluation of the expression  (\ref{Zk}) by
performing the perturbation theory expansion in coupling constants $u$, $v$.
The terms in this expansion can be presented diagrammatically
as shown in figure~\ref{fig:1}. The first diagram describes the zeroth order
approach corresponding to the idealized Gaussian chain without any interaction between the
 monomers. Solid line on the diagrams presents the
polymer chain, the dashed line describes the excluded volume
interaction between monomers governed by the coupling $u$,
and the wavy line presents  the attractive interaction caused by the presence
of the impurities governed by the coupling $v$. Let us consider the expressions
corresponding to the second and third diagram:
\begin{eqnarray}
&&D2=-u\int \rd^{d} {\vec q}\int_0^S
\rd s'\int^{s'}_{0}\rd s''{\rm e}^{-\frac{q^2}{2}(s''-s')} {\rm e}^{-\frac{k^2}{2}(S-s''+s')}, \nonumber\\
&&D3=v\int \rd^{d-\varepsilon_d}
{\vec q}\int_0^S \rd s'\int^{s'}_{0}\rd s''{\rm e}^{-\frac{q^2}{2}(s''-s')} {\rm e}^{-\frac{k^2}{2}(S-s''+s')}.
\end{eqnarray}
It is necessary to point out that in the expression for $D3$, the integration
is performed only in subspace $d-\varepsilon_d$ due to the fact that
the interaction $v$ acts only in this subspace.
Using the Poisson formula to integrate over the wave vector $\vec{q}$ we receive:
\begin{eqnarray}
&&D2=-u\frac{1}{(2\pi)^{d/2}}\int_0^S
\rd s'\int^{s'}_{0}\rd s''(s''-s')^{\frac{d}{2}}\,{\rm e}^{-\frac{k^2}{2}(S-s''+s')},
\nonumber\\
&&D3=v\frac{1}{(2\pi)^{(d-\varepsilon_d)/2}} \int_0^S
\rd s'\int^{s'}_{0}\rd s''(s''-s')^{\frac{d-\varepsilon_d}{2}} {\rm e}^{-\frac{k^2}{2}(S-s''+s')}.
\end{eqnarray}

Expanding the exponents over $\vec{k}$ and then integrating
over areas we finally receive:
\begin{eqnarray}
&&D2=-z_u\frac{1}{\left(1-\frac{d}{2}\right)\left(2-\frac{d}{2}\right)}+z_u \frac{k^2
S}{2}\frac{2}{\left(1-\frac{d}{2}\right)\left(2-\frac{d}{2}\right)\left(3-\frac{d}{2}\right)}\,, \nonumber\\
&&D3=z_v\frac{1}{\left(1-\frac{d-\varepsilon_d}{2}\right)
\left(2-\frac{d-\varepsilon_d}{2}\right)}-z_v \frac{k_{d-\varepsilon_d}^2
S}{2}\frac{2}{\left(1-\frac{d-\varepsilon_d}{2}\right)
\left(2-\frac{d-\varepsilon_d}{2}\right)\left(3-\frac{d-\varepsilon_d}{2}\right)}\,,
\end{eqnarray}
where $z_u= u (2\pi)^{-d/2}
S^{2-d/2}$ and $z_v=v (2\pi)^{-(d-\varepsilon_d)/2}
S^{2-(d-\varepsilon_d)/2}$ are dimensionless coupling constants.

Collecting all contributions from the considered diagrams one
receives an expression for the  partition function of the model by keeping terms that
do not depend on $\vec{k}$:
\[
{\overline {Z(S)}}=1-\frac{z_u
}{\left(1-\frac{d}{2}\right)\left(2-\frac{d}{2}\right)}
-\frac{z_v}{\left(1-\frac{d-\varepsilon_d}{2}\right)\left(2-\frac{d-\varepsilon_d}{2}\right)}\,.
\]
The
expressions for the components of the end-to-end distance of the polymer chain can be estimated using the identities:
\[
{\overline { \langle R^2_{d-\varepsilon_d} \rangle}}=-2\frac{1}{Z(S)}\left[\frac{\rd}{\rd\vec{k}_{d-\varepsilon_d}}\tilde{Z}(\vec{k},S)\right]_{\vec{k}=0}, \qquad
{\overline {\langle R^2_{\varepsilon_d} \rangle}}=-2\frac{1}{Z(S)}\left[\frac{\rd}{\rd\vec{k}_{\varepsilon_d}}\tilde{Z}(\vec{k},S)\right]_{\vec{k}=0}.
\]
We distinguish between
the components in subspaces $\varepsilon_d$ and $d-\varepsilon_d$,
corresponding to components of the end-to-end distance in directions
parallel and perpendicular to extended defects:
\begin{align}
{\overline {\langle R^2_{d-\varepsilon_d} \rangle}}&=S(d-\varepsilon_d)\left[1+\frac{z_u}{\left(2-\frac{d}{2}\right)
\left(3-\frac{d}{2}\right)}-\frac{z_v}{\left(2-\frac{d-\varepsilon_d}{2}\right)
\left(3-\frac{d-\varepsilon_d}{2}\right)}\right],\label{RRpar}\\
{\overline {\langle R^2_{\varepsilon_d} \rangle}}&=S\varepsilon_d
\left[1+\frac{z_u}{\left(2-\frac{d}{2}\right)\left(3-\frac{d}{2}\right)}\right].\label{RRperp}
\end{align}
References (\ref{RRpar}) and (\ref{RRperp}) confirm the existence of two characteristic lengths for polymers in
aniso\-tropic environments.
The presence of extended defects makes the polymer radius shrink
in transverse direction  due to the attractive interactions between monomers governed by the coupling $v$,
whereas in parallel direction,
the increase of the effect of repulsive interactions (as consequence of the increase of monomer density)
is responsible for the elongation of  the polymer chains.

\begin{figure}[!t]
\begin{center}
\includegraphics[width=85mm]{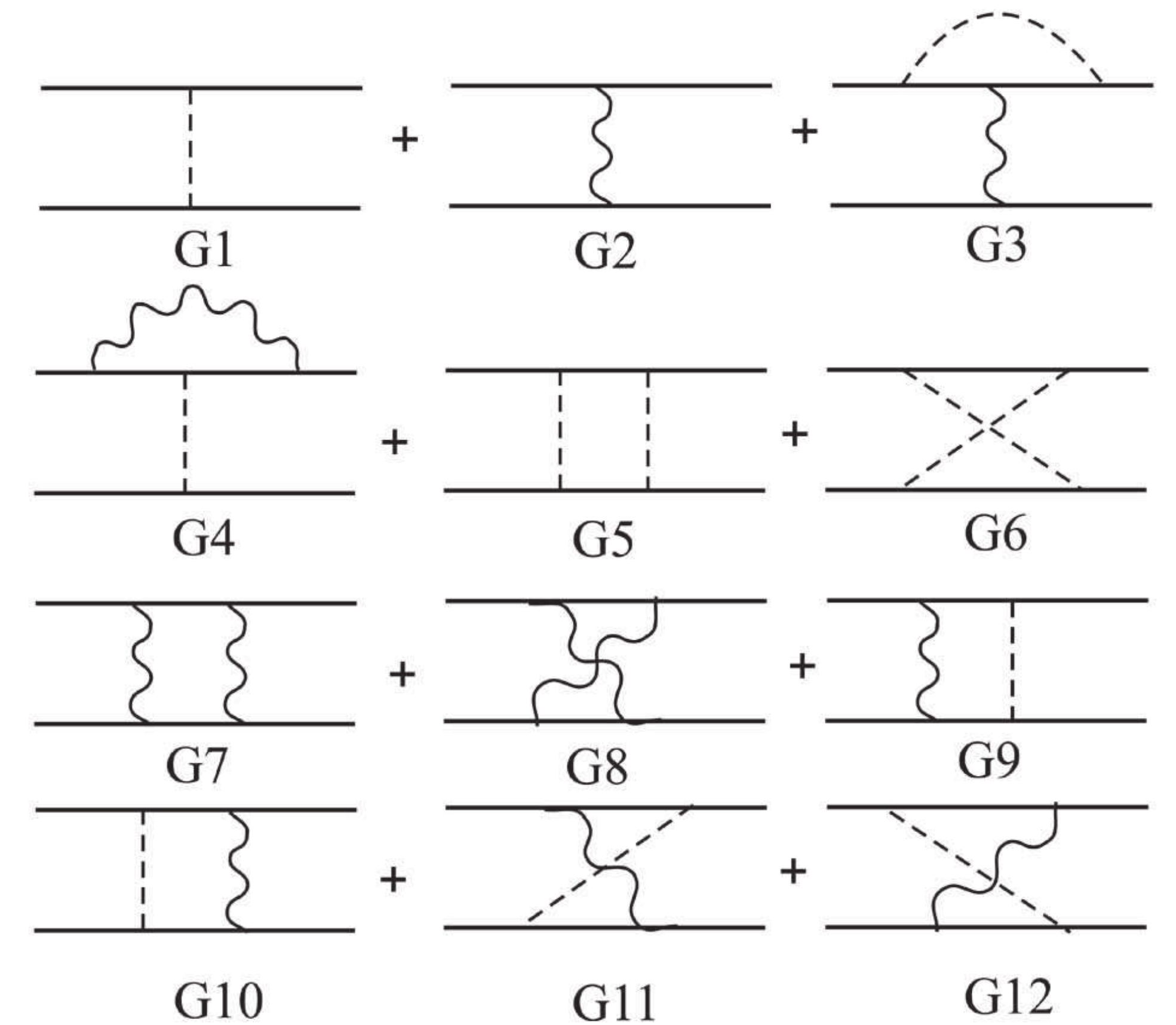}
\end{center}
\caption{Diagramatic presentation of the contributions into the partition
function $Z(S,S)$ of two interacting polymer chains up to the second order of expansion in the coupling constants.} \label{fig:2}
\end{figure}

Calculating contributions to the partition function of two interacting
polymer chains one may use a diagrammatic representation  (see figure~\ref{fig:2}).
Note that only those diagrams are taken into account which contain
at least one interaction line.
The first few diagrams, those with one interaction acting between two
polymers ($G1-G4$), can be gathered and presented as
$-uS^2Z(S)^2-vS^2Z(S)^2$.
Performing the dimensional analysis of the contributions, produced by
different diagrams, we find two distinct classes of graphs. The
first class of graphs produces terms which behave like
$[S]^{\frac{4-d}{2}}$, the sum of all such terms gives
contributions into the function denoted by $Z_u(S,S)$. The  diagrams of the second
class  behave like $[S]^{\frac{4-d+\varepsilon_d}{2}}$
and thus give contributions into the function $Z_v(S,S)$.
As a result, the ``two polymer function'' can be presented in the
form: $
Z(S,S)=Z_u(S,S)+Z_v(S,S),
$
where $Z_u(S,S)$ and $Z_v(S,S)$ are given by the expressions:
\begin{eqnarray}
Z_u(S,S)&=&-uS^2\Bigg[1+2\frac{z_u}{\left(1-\frac{d}{2}\right)\left(2-\frac{d}{2}\right)}
-2\frac{z_v}{\left(1-\frac{d-\varepsilon_d}{2}\right)\left(2-\frac{d-\varepsilon_d}{2}\right)}
\nonumber\\
&&+2z_u\frac{2^{4-d/2}-10+d}{\left(1-\frac{d}{2}\right)\left(2-\frac{d}{2}\right)
\left(3-\frac{d}{2}\right)\left(4-\frac{d}{2}\right)}
-4z_v\frac{2^{4-(d-\varepsilon_d)/2}-10+d}{\left(1-\frac{d-\varepsilon_d}{2}\right)\left(2-
\frac{d-\varepsilon_d}{2}\right)\left(3-\frac{d-\varepsilon_d}{2}\right)
\left(4-\frac{d-\varepsilon_d}{2}\right)}\Bigg],\nonumber
\end{eqnarray}
\vspace{-5mm}
\begin{eqnarray}
Z_v(S,S)&=&vS^2\Bigg[1+2\frac{z_u}{\left(1-\frac{d}{2}\right)\left(2-\frac{d}{2}\right)}
-2\frac{z_v}{\left(1-\frac{d-\varepsilon_d}{2}\right)\left(2-\frac{d-\varepsilon_d}{2}\right)}
\nonumber\\
&&-2z_v\frac{2^{4-(d-\varepsilon_d)/2}-10+d}{\left(1-\frac{d-\varepsilon_d}{2}\right)
\left(2-\frac{d-\varepsilon_d}{2}\right)\left(3-\frac{d-\varepsilon_d}{2}\right)
\left(4-\frac{d-\varepsilon_d}{2}\right)}\Bigg].
\end{eqnarray}

The swelling factor in our model reads:
\begin{align}
\chi^{0}&=\frac{R^2}{S
d}=\frac{\varepsilon_d}{d}R^2_{\varepsilon_d}+\frac{d-\varepsilon_d}{d}R^2_{d-\varepsilon_d}
\nonumber\\[1ex]
&=\left[1+\frac{z_u}{\left(2-\frac{d}{2}\right)\left(3-\frac{d}{2}\right)}
+\frac{d-\varepsilon_d}{d}\frac{z_v}{\left(2-\frac{d-\varepsilon_d}{2}\right)
\left(3-\frac{d-\varepsilon_d}{2}\right)}\right].
\end{align}
The renormalized coupling constants can be presented in the form:
\begin{align*}
g_u&=\chi_1^{-4}\chi_0^{-2+\varepsilon/2}Z_u(S,S),\\
g_v&=\chi_1^{-4}\chi_0^{-2+\delta/2}Z_v(S,S).
\end{align*}
The corresponding  flow equations read:
\begin{align*}
W[g_u]&=\varepsilon g_u-8 g_u^2+12g_u g_v\,,\\
W[g_v]&=-\delta g_v-8g_v^2+4 g_u g_v\,,
\end{align*}
here, $\varepsilon=4-d$, $\delta=\varepsilon+\varepsilon_d$. The coordinates of fixed points can be found as common zeros of functions $W[g_u]$,$W[g_v]$:
\begin{align}
&g^*_u=0,& &g^*_v=0, \label{gg}\\
&g^*_u=\varepsilon/8,&  &g^*_v=0, \label{pure}\\
&g^*_u=0,& &g^*_v=-\delta/8, \label{rw}\\
&g^*_u=\varepsilon/2-(3/4)\delta,&  &g^*_v=\varepsilon/4-\delta/2
\label{exten}.
\end{align}
The first fixed point describes  the case  of an idealized
Gaussian chain without any interactions between monomers. Expression (\ref{pure}) corresponds to the case of a polymer chain
with short-range excluded
volume interactions in a pure solvent. The fixed points (\ref{rw}) and (\ref{exten})
 describe, correspondingly,
the Gaussian chain and the chain with excluded volume effect in the anisotropic environment.
However, since both of them are associated with attractive interactions between monomers
due to the presence of defects,
these fixed points appear to be unstable in
the physical region of the parameters ($\varepsilon>0$ and $\varepsilon_d>0$)
and thus cannot provide estimates of
scaling exponents.
 Note that a similar problem of the absence of
stable and physically accessible fixed points  also exists in the
case of uncorrelated point-like impurities \cite{Chakrabarti81}. The latter was solved by absorbing the interaction with disorder into the
excluded volume interaction due to a special symmetry \cite{Kim83}.
However, this does not work in the present case of extended defects.

\section{Conclusions}

 We analyzed the influence of anisotropy of the environment
 caused by the presence of impurities  correlated  in $\varepsilon_d$ dimensions,
 on conformational size and shape characteristics of long flexible polymer chains.
 The integer values of $\varepsilon_d$ have direct physical interpretation and
 describe extended defects, e.g., in the form of lines or planes of parallel orientation ($\varepsilon_d=1$ or $2$, correspondingly).
 In this case, it is obvious that one should distinguish between two characteristic length scales, in directions parallel and perpendicular to
 such extended defects.
 Non-integer values of $\varepsilon_d$ may correspond to complex defects of fractal nature.

Applying the numerical simulations based on the model of self-avoiding random walks on
a regular cubic lattice,  we considered three
cases: impurities in the form of parallel lines ($\varepsilon_d=1$),
fractal-like structures with $0<\varepsilon_d<1$ (which can be treated as partially penetrable lines)
and fractal-like structures with $1<\varepsilon_d<2$ (partially penetrable planes).
In the first case, we found that  components of the effective linear size of polymer chain, that are either parallel or perpendicular
to the lines of impurities, behave differently and their scaling is governed by two distinct scaling
exponents $\nu_{\|}$ and $\nu_{\bot}$ (see equation (\ref{R1})). The exponent governing the scaling of a parallel component
of the end-to-end distance gradually reaches the maximal value of 1 with increasing of
concentration of defects, while $\nu_{\bot}$  gradually tends to zero.
Analyzing the influence of disorder in the form of partially penetrable lines on scaling properties of
polymers, we again found the existence of two distinct
exponents $\nu_{\|}$ and  $\nu_{\bot}$  (see figure~\ref{fig:11}~(a)). This tendency (and thus the anisotropy)
surprisingly persists  even at high  probability of the  polymer chain to penetrate through such ``line'' (which corresponds to
$\varepsilon_d$ close to $0$).
Considering  structural defects in the form of
partially penetrable planes of parallel orientation (see figure~\ref{fig:0}~(c)),  we found that
the exponent $\nu_{\|}$  gradually changes from the value found earlier for the three-dimensional pure lattice to that in two dimensions
with growing concentration of impurity planes. This can be treated
as a crossover to a restricted geometry regime of the polymer confined
between two homogeneous planes. The exponent $\nu_{\bot}$ gradually tends to
zero.

Our analytical studies were performed within the frames of the direct polymer renormalization approach
using the double $\varepsilon$, $\varepsilon+\varepsilon_d$ expansion.
In particular, we found expressions for
the components of the end-to-end distance of polymer chain (\ref{RRpar}), (\ref{RRperp}). The presence of extended defects makes the polymer radius shrink
in transverse direction  due to attractive interactions between monomers governed by the coupling $v$,
whereas in parallel direction
the increase of the effect of repulsive interactions (as a consequence of the increase of monomer density)
is responsible for the elongation of the polymer chain. We conclude
  that the presence of extended defects correlated in
 $\varepsilon_d$ dimensions makes the polymer chain elongated in the direction parallel to these extended impurities,
 which confirms the existence of two characteristic lengths for polymers	 in
anisotropic environments.

\section*{Acknowledgements}
This work was supported in part by the
FP7 EU IRSES projects N269139  ``Dynamics and Cooperative Phenomena in Complex
Physical and Biological Media'' and N295302 ``Statistical Physics in Diverse Realizations''.

\clearpage

\ukrainianpart
\title{Конформаційні властивості полімерів в анізотропних середовищах}
\author{К. Гайдуківська, В. Блавацька}
\address{
Інститут фізики конденсованих систем НАН України,  вул. І. Свєнціцького, 1, 79011 Львів, Україна
}

\makeukrtitle

\begin{abstract}
\tolerance=3000%

Проаналізовано конформаційні властивості полімерних макромолекул у розчинах
в присутності протяжних структурних домішок (фрактальної) вимірності $\varepsilon_d$, що спричиняють просторову анізотропію.
Застосовуючи збіднено-збагачений алгоритм Розенблюта (PERM),
  отримано чисельні оцінки для скейлінгових показників та універсальних характеристик форми
 полімерів у таких середовищах у широкому спектрі $0<\varepsilon_d<2$ при вимірності простору $d=3$.
  Аналітичний опис моделі розвинено в рамках підходу прямого полімерного перенормування де Клуазо.
  Як чисельні, так і аналітичні дослідження підтверджують
 існування двох характеристичних масштабів довжини полімерного ланцюжка
 в паралельному та перпендикулярному напрямках до протяжних дефектів.

\keywords полімери, заморожений безлад, скейлінг, ренормалізаційна група, чисельні симуляції
\end{abstract}

\end{document}